\documentclass[aps,prl,twocolumn]{revtex4}

\usepackage{rotating}
\usepackage{graphicx}

\begin{document}

\title{Collapse-driven spatiotemporal dynamics of filament formation}
\author{Miguel A. Porras$^1$, Alberto Parola$^2$, Daniele Faccio$^2$, A. Couairon$^3$, Paolo Di Trapani$^{2,4}$}
\affiliation{$^1$Departamento de F\'{\i}sica Aplicada, Universidad Polit\'ecnica de Madrid, Rios Rosas 21, ES-28003, Spain \\
             $^2$INFM and Department of Physics, University of Insubria, Via Valleggio 11, IT-22100
             Como, Italy\\
             $^3$Centre de Physique Th\'eorique, CNRS, \'Ecole Polytechnique, F-91128, Palaiseau, France\\
             $^4$Department of Quantum Electronics, Vilnius University, Sauletekio 9, LT 01222, Vilnius, Lithuania}
\begin{abstract}
The transition from spatial to spatiotemporal dynamics in Kerr-driven beam collapse is modelled as the instability of the Townes
profile. Coupled axial and conical radiation, temporal splitting and X waves appear as the effect of Y-shaped unstable modes,
whose growth is experimentally detected.
\end{abstract}

\maketitle

From light filaments to Bose Einstein condensates (BEC), from plasma instabilities to hydrodynamical or optical shocks, many
nonlinear wave processes lead to self-compression with catastrophic increase in peak intensity, followed by relaxation to a
linear state. Universal properties of the compression dynamic were discovered by analyzing the blowup of self-focusing solutions
to the nonlinear Schr\"odinger equation (NSE) \cite{FIBICH}, described as a self-similar collapse to a smooth and symmetric
ground state \cite{RYPDAL}. Though wave-packets evolve in the three-dimensional (3D) space, uneven initial conditions and
compression rates typically lead to symmetry breaking, featured by a transient compression towards a sub-dimensional eigenstate.
In light-beam self-focusing, self-similar collapse to the Townes profile (TP), or ground state of the NSE in two spatial
dimensions \cite{CHIAO}, has been recently observed \cite{GAETA}. As to the subsequent expansion, 3D (space-time) X-waves were
shown to capture the apparent stationarity, the pulse-splitting and the conical emission (CE) in the filamentation regime
\cite{KOLESIK,FACCIO2}. Up to date, however, the mechanisms that support the transition from the quasi-2D, self-similar collapse
to the 3D relaxation regime remain unrevealed.

A possible approach to this problem requires retaining all highly nonlinear effects relevant to this transition phase in a
suitably ``dressed" NSE. Here we show that the collapse-driven ``morphology" transition from the dominant 2D to the 3D dynamics
can be described in the frame of the bare 3D NSE model by the 3D (space-time) instability of the 2D TP eigensate. Up to date,
transverse instability was investigated only for the 1D NSE eigenstate \cite{KUZNETSOV,RYPDAL2}, which however does not support
collapse. Our analysis directly applies to light filaments generated by fs pulses in normally dispersive Kerr media, but can be
generalized straightforwardly to the collapse in anomalous dispersion, the spatiotemporal (ST) instability of $\chi^{(2)}$
solitons and the collapse of sub-dimensional localized structures in plasmas and BEC. The results capture the key steps of the
entire filament evolution from the input Gaussian to the final X waves, and explain the coupling between axial and CE. These
were regarded as two independent processes, until recent numerical investigations outlined their simultaneous appearance
\cite{BRAGHERI}. Here the coupling emerges as an intrinsic feature of the instability of the TP, whose signature is a couple of
Y-shaped unstable modes. For the first time, a theory is provided that links filamentation, CE, pulse-splitting and
supercontinuum generation to the singular behavior at collapse.

\begin{figure}
\includegraphics[width=3.8cm]{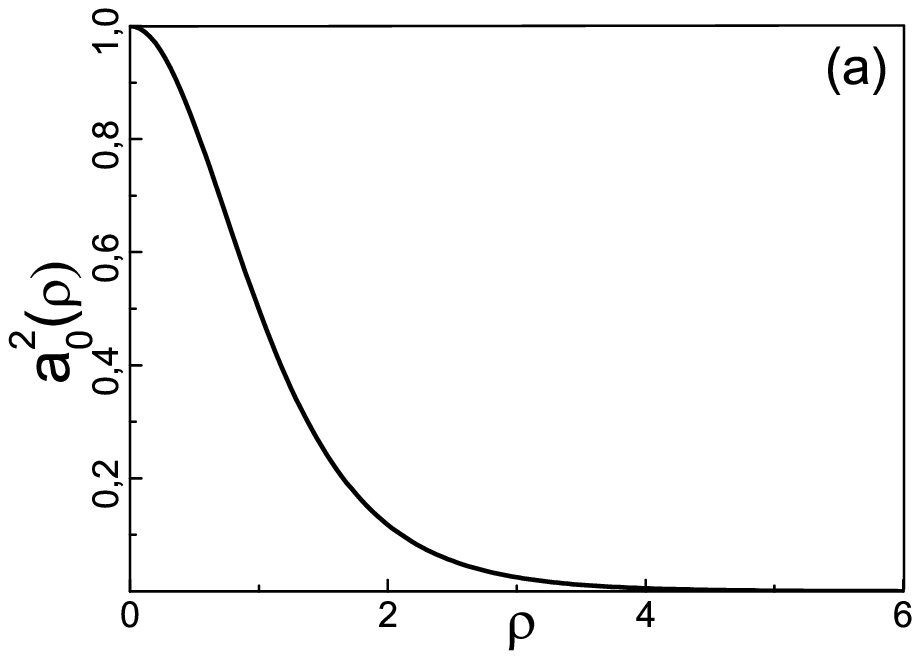}\includegraphics[width=3.75cm]{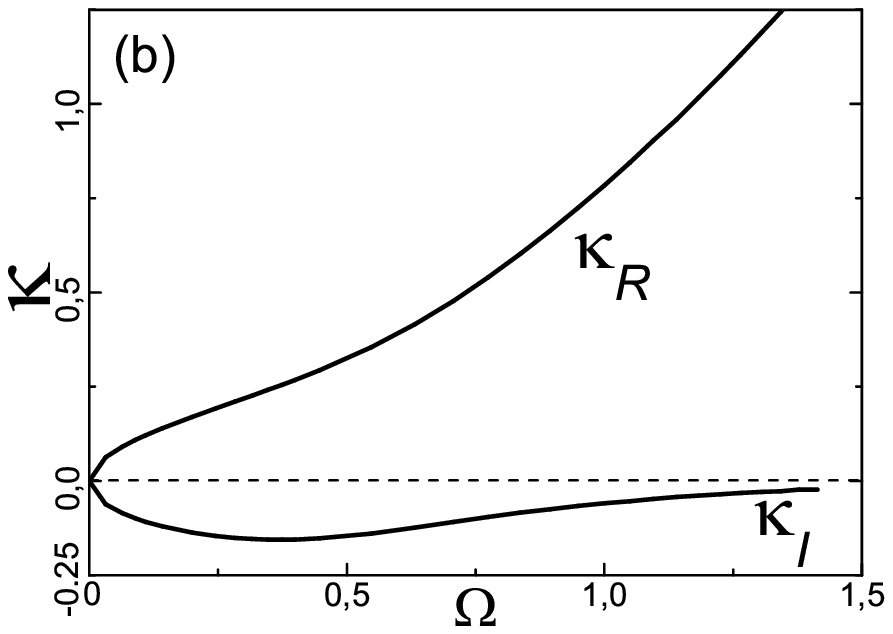}
\caption{\label{fig1} (a) Normalized TP. (b) Real and imaginary parts of the relevant complex eigenvalue.}
\end{figure}

Consider the cubic, dimensionless NSE
\begin{equation}\label{NSE}
\partial_\zeta A = \frac{i}{2} \Delta_{\xi,\eta} A -\frac{i}{2} \partial^2_\tau A+ i|A|^2A\, ,
\end{equation}
with $\Delta_{\xi,\eta} =\partial^2_{\xi}+\partial^2_{\eta}$, for a function $A$ that depends on the spatial coordinates
$(\xi,\eta)$ and the temporal coordinate $\tau$. The ground state of the NSE in two spatial dimensions is expressed by
$A=a_0(\rho)\exp(i\alpha\zeta)$ [$\rho=(\xi^2+\eta^2)^{1/2}$], where $a_0(\rho)$ is the TP \cite{CHIAO} [Fig. \ref{fig1}(a)],
and $\alpha\simeq 0.2055$ for the normalization $a_0(0)=1$.

Written in the variables $z=(n_0/k_0n_2I)\zeta$, $(x,y)=(n_0/k_0^2n_2I)^{1/2}(\xi,\eta)$,
$t-k'_0z=(n_0k_0^{\prime\prime}/k_0n_2I)^{1/2}\tau$, $\tilde A=I^{1/2} A$, Eq. (\ref{NSE}) is a standard model for the
propagation of an optical wave packet $E=\tilde A(x,y,z,t)\exp(-i\omega_0t+ik_0z)$ subject to diffraction, normal group velocity
dispersion (GVD) ($k_0^{\prime\prime}>0$) and self-focusing nonlinearity ($n_2>0$) leading to collapse. The ground state
describes a monochromatic light beam with $z$-independent transverse amplitude profile. In the above relations, $\omega_0$ is
the carrier frequency of the wave packet, $I$ is a characteristic intensity, $n_2$ the nonlinear refraction index of the medium,
and $k(\omega)=n(\omega)\omega/c$ the propagation constant, with $n(\omega)$ the refractive index and $c$ the speed of light in
vacuum. Prime signs stand for differentiation with respect to $\omega$, and subscripts 0 for evaluation at $\omega_0$.

Following a standard MI analysis, the perturbed TP
\begin{equation}\label{PERTURBATION}
A=\{a_0(\rho)+ \sigma[u(\rho)e^{-i\Omega\tau+i\kappa\zeta}
                                      +v^\star(\rho)e^{i\Omega\tau-i\kappa^\star\zeta}]\}e^{i\alpha\zeta}\, ,
\end{equation}
with $\sigma\ll 1$, is considered. The ST perturbation will grow exponentially with $\zeta$ if $\kappa_I<0$ ($\kappa_I\equiv
{\rm Im} \kappa$). Opposite frequency shifts $\Omega$ and $-\Omega$ characterize the components $u$ and $v^\star$. For
simplicity, only cylindrically symmetric perturbations are considered. Introducing (\ref{PERTURBATION}) into (\ref{NSE}), and
keeping terms up to the first-order in $\sigma$, one gets the differential eigenvalue problem
\begin{equation}\label{PROBLEM}
\left( \begin{array}{cc} H & a_0^2 \\ -a_0^2 & -H \end{array}\right)\left(\begin{array}{c} u \\ v\end{array}\right) =
                \kappa\left(\begin{array}{c} u \\ v\end{array}\right)\, ,
\end{equation}
where $H= \frac{1}{2}[d^2/d\rho^2 + (1/\rho)d/d\rho + \Omega^2]-\alpha +2a_0^2(\rho)$, with boundary conditions $du(0)/d\rho
=dv(0)/d\rho=0$. Given a perturbation frequency $\Omega$, if the problem admits at least one nontrivial solution $(u,v)$ with an
eigenvalue $\kappa$ with negative imaginary part, then the TP is unstable under perturbations at that frequency. The problem
(\ref{PROBLEM}) has to be solved numerically. To simplify the procedure, note first that if $(\kappa,u,v)$ is a solution of
(\ref{PROBLEM}) at $\Omega$, then $(\kappa^\star,u^\star,v^\star)$, $(-\kappa, v,u)$ and $(-\kappa^\star, v^\star,u^\star)$ are
solutions too, whose eigenvalues are all reflections of $\kappa$ about the real and imaginary axes in the complex
$\kappa$-plane. Then, if $\kappa$ is complex, only two of these solutions represent unstable perturbations. Following the
convention that $\kappa_R\ge 0$ ($\kappa_R\equiv {\rm Re}\kappa$) and $\kappa_I<0$, these two solutions are $(\kappa,u,v)$ and
$(-\kappa^\star, v^\star,u^\star)$, that yield the two independent, unstable and {\em physically distinguishable} perturbations
\begin{eqnarray}\label{YY}
p_\kappa(\rho,\Omega)\!\!\!&=&\!\!\! \left[u(\rho) e^{-i\Omega\tau+i \kappa_R\zeta}
                    \!\!+\!v^\star\!(\rho) e^{i\Omega\tau-i\kappa_R\zeta}\right]e^{\!-\!\kappa_I\zeta} \,, \nonumber \\
p_{\!-\!\kappa^\star}(\rho,\Omega)\!\!\!&=&\!\!\! \left[v^\star\!(\rho) e^{-i\Omega\tau-i\kappa_R\zeta}
                                      \!\!+\!u(\rho) e^{i\Omega\tau+i\kappa_R\zeta}\right]e^{\!-\!\kappa_I\zeta},
\end{eqnarray}
since $u$ and $v^\star$ are generally different and oppositely frequency shifted in $p_\kappa$ and $p_{-\kappa^\star}$. Also,
any solution of (\ref{PROBLEM}) with $\Omega$ is also solution for $-\Omega$. In particular, the unstable solutions
$(\kappa,u,v)$ and $(-\kappa^\star, v^\star,u^\star)$ for the perturbation frequency $-\Omega$ are seen from
(\ref{PERTURBATION}) to represent the same but permuted perturbations $p_{-\kappa^\star}$ and $p_\kappa$. In short, we can
restrict the analysis of (\ref{PROBLEM}) to $\Omega\ge 0$, and count only eigenvalues with $\kappa_R\ge 0$. For each of these
with $\kappa_I<0$, two independent, unstable perturbations (\ref{YY}) may grow.

The problem (\ref{PROBLEM}) has been solved numerically for each $\Omega\ge 0$ after discretization of the differential
operators on a radial ($\rho$) grid of finite size much larger than the Townes range $d\sim 1.5$. Increasing grid points and
size allowed to control the accuracy of the results. For $\Omega=0$, no complex eigenvalue is found, as expected from the
marginal instability of the TP under pure spatial perturbations \cite{RYPDAL2}. For each $\Omega>0$, only one, isolated
eigenvalue $\kappa$ with $\kappa_I<0$ appears, whose real and imaginary parts are shown in Fig. \ref{fig1}(b). The ground state
$a_0(\rho) \exp (i \alpha \rho)$ is then retrieved to be modulationally unstable under ST perturbations \cite{SHEN}. The gain
$-\kappa_I$ is limited to $\Omega \lesssim 1.5$ but without an abrupt cut-off.

For the eigenmodes $u$ and $v$ associated with the complex eigenvalue $\kappa$, some relevant features can be inferred without
resorting to numerical calculation. Neglecting terms with $a_0^2$ in (\ref{PROBLEM}) and in $H$ for $\rho\rightarrow\infty$, the
decoupled equations $\Delta_{\xi,\eta} u = -[\Omega^2 -2\alpha-2\kappa]u$ and $\Delta_{\xi,\eta} v=-[\Omega^2 -2\alpha+2\kappa]
v$ are obtained. Their bounded solutions for $\rho\rightarrow\infty$ are $u\propto H_0^{(1)}\left[(Q_u+i\Gamma_u)\rho\right]$
and $v\propto H_0^{(1)}\left[(Q_v+i\Gamma_v)\rho\right]$, where $H_0^{(1)}$ is the H\"ankel function of first class and zero
order, and the real quantities $Q_{u,v}$ and $\Gamma_{u,v}$ are defined by
\begin{equation}\label{QGAMMA}
Q_u\!+\!i\Gamma_u\!=\!\! \sqrt{\Omega^2\! -\!2\alpha\!-\!2\kappa},\,\, Q_v\!+\!i\Gamma_v\!=\!\! \sqrt{\Omega^2\!
-\!2\alpha\!+\!2\kappa},
\end{equation}
with the convention of taking square roots such that $\Gamma_{u,v}\ge 0$. Using that $H^{(1)}_0(s)\sim \sqrt{(2/\pi
s)}\exp{[i(s-\pi/4)]}$ for large $|s|$, ignoring constant factors and algebraical decay, the expressions $u(\rho)\sim
\exp{(-\Gamma_u\rho + iQ_u\rho)}$ and $v(\rho)\sim \exp{(-\Gamma_v\rho + iQ_v\rho)}$ are found to describe the dominant behavior
of the unstable eigenmodes for large $\rho$. Thus, if their decay rates are small ($\Gamma_{u,v}\ll 1$), $u$ and $v$ are
expected to feature radial oscillations of transverse wave numbers $Q_{u,v}$ beyond the TP range.

\begin{figure}
\includegraphics[width=4.2cm]{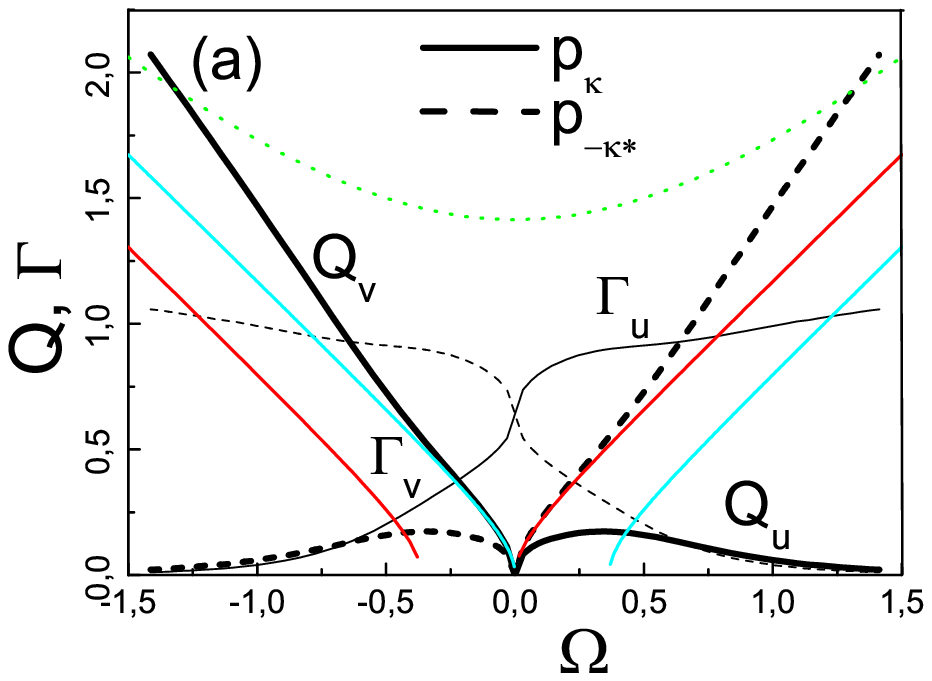}\includegraphics[width=4.2cm]{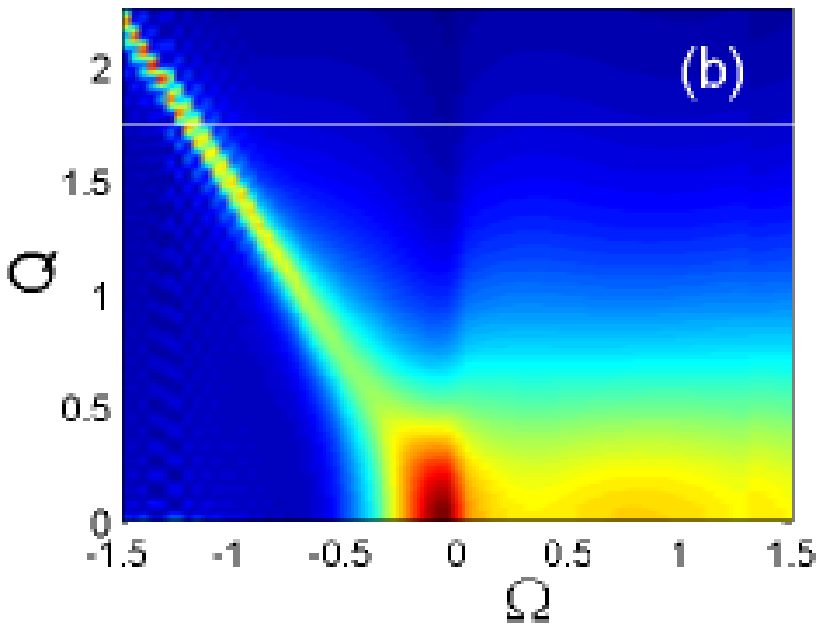}
\caption{\label{fig2} (a) Thick black curves: Characterization of the ST spectrum of the unstable perturbations $p_\kappa$ and
$p_{-\kappa^\star}$ by means of the transverse frequencies $Q_{u,v}$. Thin black curves: Exponential radial decay rates
$\Gamma_{u,v}$. Green curve: maximum gain curve of the plane wave. Red and blue curves: double-X spectrum. (b) Modulus of the ST
spectrum $\hat p_\kappa(Q,\Omega)$ of $p_\kappa$. The amplitudes for different $\Omega$ are arbitrarily chosen so that the
energy $\int_0^\infty |\hat p_\kappa|^2QdQ$ is independent of $\Omega$.}
\end{figure}

The physical quantities featuring the collapsing dynamics are the growing perturbations $p_\kappa$ and $p_{-\kappa^\star}$
composed of $u$ and $v$. In particular, the ST spectrum $Q$--$\Omega$ of the unstable perturbations at a certain propagation
distance $\zeta$ will be generally expressed by a superposition of the type $\sigma_\kappa(\Omega)\hat p_\kappa(Q,\Omega) +
\sigma_{-\kappa^\star}(\Omega) \hat p_{-\kappa^\star}(Q,\Omega)$, where $\hat p_{\kappa}(Q,\Omega)$ and $\hat
p_{-\kappa^\star}(Q,\Omega)$ are the spatial Fourier transforms of $p_{\kappa}(\rho,\Omega)$ and
$p_{-\kappa^\star}(\rho,\Omega)$, and $\sigma_\kappa(\Omega)$ and $\sigma_{-\kappa^\star}(\Omega)$ are the seeds of each
unstable perturbation at each frequency. In experiments, measurement of the ST spectrum $Q$--$\Omega$ (in practice, the
angularly-resolved spectrum, displaying angles and wavelengths) of a light filament is a powerful diagnostic in which axial and
CE can be visualized at once \cite{FACCIO}. For the TP, the structure of the $Q$--$\Omega$ spectrum of instability can be
inferred from the above asymptotic analysis. The spectrum $\hat p_\kappa(Q,\Omega)$ is composed of $\hat u(Q,\Omega)$ at
positive frequency shift $\Omega$ and expectedly peaked at $Q_u$, and of $\hat v^\star(Q,\Omega)$ at negative frequency shift
$-\Omega$ and peaked at $Q_v$ (if the respective decay rates $\Gamma_{u,v}$ are not large). Plotting $Q_u$ and $Q_v$ given by
(\ref{QGAMMA}) versus frequency in their respective ranges, we obtain the black, thick solid curve of Fig. \ref{fig2}(a). This
curve is expected to locate the regions in the $Q$--$\Omega$ plane where $\hat p_\kappa(Q,\Omega)$ takes its highest values.
Fig. \ref{fig2}(a) shows also the values of $\Gamma_{u}$ and $\Gamma_{v}$ (black, thin solid curve). Since exponential
localization of $v$ is weak ($\Gamma_v\ll 1$), the branch $Q_v$, fitting approximately the linear relation $Q_v\simeq
-\sqrt{2}\,\Omega$, is expected to define the locus of maxima of $\hat v(Q,\Omega)$, and hence $\hat p_{\kappa}(Q,\Omega)$ at
down-shifted frequencies to describe an actual off-axis, CE. Instead, localization of $u$ is similar to that of the TP
($\Gamma_u\sim 1$) and $Q_u$ close to zero, which suggests to associate $\hat p_{\kappa}(Q,\Omega)$ at up-shifted frequencies
with an axial emission. Similar analysis holds for $\hat p_{-\kappa^\star}(Q,\Omega)$, CE being now associated with up-shifted
frequencies, and axial emission with down-shifted frequencies [see Fig. \ref{fig2}(a), black, dashed curves]. This
characterization of the instability spectrum is confirmed by numerical evaluation of $u$ and $v$. Fig. \ref{fig2}(b) shows $\hat
p_\kappa(Q,\Omega)$ with the expected features. Its reflection about $\Omega=0$ yields $\hat p_{-\kappa^\star}(Q,\Omega)$. The
instability of the TP then reveals the link between axial and CE. In this view, the basic ingredients of the ST spectrum of a
light filament are not the axial continuum and the X-shaped CE, but two Y-shaped objects, each one coupling off-axis emission to
axial emission on opposite frequency bands.

Let us compare these results with those for the ST instability of the plane wave solution $A=\exp(i\zeta)$ to the NSE
\cite{LIOU}. The problem (\ref{PROBLEM}) holds for this case if the amplitude $a_0$ and the nonlinear wave vector shift $\alpha$
are replaced by unity, and presents the same symmetries. The eigenmodes $u$ and $v$ associated with the most unstable
perturbations $p_\kappa$ and $p_{-\kappa^\star}$ at each frequency $\Omega$ are two plane waves with identical transverse wave
numbers $Q_u=Q_v=\sqrt{\Omega^2+2}$. The gain $-\kappa_I=1$ is independent of $\Omega$ and hence unbounded. The spectra $\hat
p_\kappa(Q,\Omega)$ and $\hat p_{-\kappa^\star}(Q,\Omega)$ are then characterized by two identical hyperbolas, depicted in Fig.
\ref{fig2}(a) as a single green curve (see Ref. \cite{LIOU}), and usually referred to as an X-like spectrum. The slope of the X
arms is unity, while that of the Y arm is $\sqrt{2}$ ($\sqrt{k_0k_0^{\prime\prime}}$ against $\sqrt{2k_0k_0^{\prime\prime}}$ in
physical units).

Degeneracy breaking from single X to double Y and gain limitation originate from the {\em transverse localization} of the TP. To
see this more intuitively, we interpret instability as originated from a Kerr-driven, degenerate four-wave-mixing (FWM)
interaction in which two intense, identical pump waves of frequency $\Omega=0$ propagating along the $\zeta$ direction, amplify
two weak, non-collinear plane waves, $u$ and $v$, at frequencies $\pm\Omega$. Each pump wave propagates with an axial wave
number shift $k_{\rm NL}$ (with respect to a plane wave of frequency $\Omega=0$) due to self-phase modulation (SPM). The axial
wave numbers of the weak plane waves are shifted by $\Omega^2/2-Q_u^2/2$ and $\Omega^2/2-Q_v^2/2$ [as obtained from the
linearized NSE (\ref{NSE})] due to material dispersion and tilt, and by $k_{{\rm NL},u}$ and $k_{{\rm NL},v}$ due to cross-phase
modulation (XPM). Axial phase matching among the interacting waves then reads as
\begin{equation}\label{MATCHING}
\Omega^2/2 - Q_u^2/2 + k_{{\rm NL},u} + \Omega^2/2 - Q_v^2/2 +  k_{{\rm NL},v} = 2 k_{\rm NL}\, .
\end{equation}
For plane wave pumps, SPM and XPM wave number shifts are $k_{\rm NL}=1$, $k_{{\rm NL},u}= k_{{\rm NL},v}= 2 k_{\rm NL} = 2$
\cite{ALFANO}. Axial and transverse phase-matching ($Q_u^2=Q_v^2$) must be strictly enforced for efficient amplification of $u$
and $v$, which lead to $Q_{u,v}=\sqrt{\Omega^2 + 2}$, i.e., to the X-like spectrum of instability of the plane wave.

Suppose, instead, that the pumps have TPs ($k_{\rm NL}=\alpha$). To account for transverse localization, a) we will assume that
plane waves collinear with the pump are preferentially amplified, and b) we take into account that efficient amplification is
also possible with small transverse phase-mismatch such that $|Q_u-Q_v|\lesssim \pi/d$ \cite{PENZKOFER}. If for instance, the
plane wave $u$ at $+\Omega$ is taken as collinear, then $Q_u\simeq 0$, and $k_{{\rm NL},u}\simeq 2k_{\rm NL}=2\alpha$. Axial
phase matching (\ref{MATCHING}) then requires $Q_v\not\simeq 0$, i.e., the plane wave $v$ at $-\Omega$ cannot be collinear. It
is then reasonable to take its XPM wave number shift as $k_{{\rm NL},v}\simeq 0$, since the beam does not remain into the
localized interaction area \cite{FACCIO2}. With these assumptions, (\ref{MATCHING}) leads to the linear relation $Q_v \simeq
-\sqrt{2}\,\Omega$, as for the Y-tail in the instability spectrum of the TP. Limitation of the gain band finds also explanation
from the maximum allowed transverse mismatch. If $Q_u\simeq 0$, then $Q_v\lesssim\pi/d$. Taking $d\simeq 1.5$ for the TP, we
obtain $Q_v\lesssim 2$, which corresponds roughly to $\Omega\lesssim 1.5$, as obtained above. Obviously, if the collinear plane
wave $u$ is taken at $-\Omega$, similar considerations lead to the reflected Y wave.

\begin{figure}
\includegraphics[width=3.5cm]{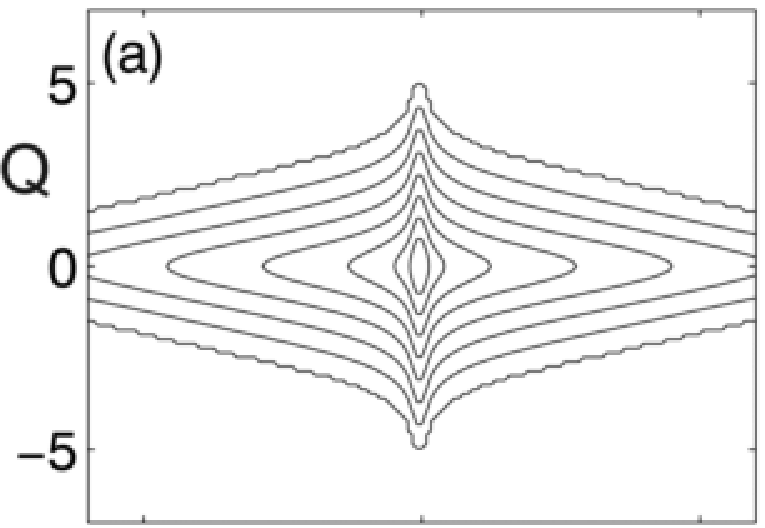}\hspace*{0.35cm}\includegraphics[width=3.8cm]{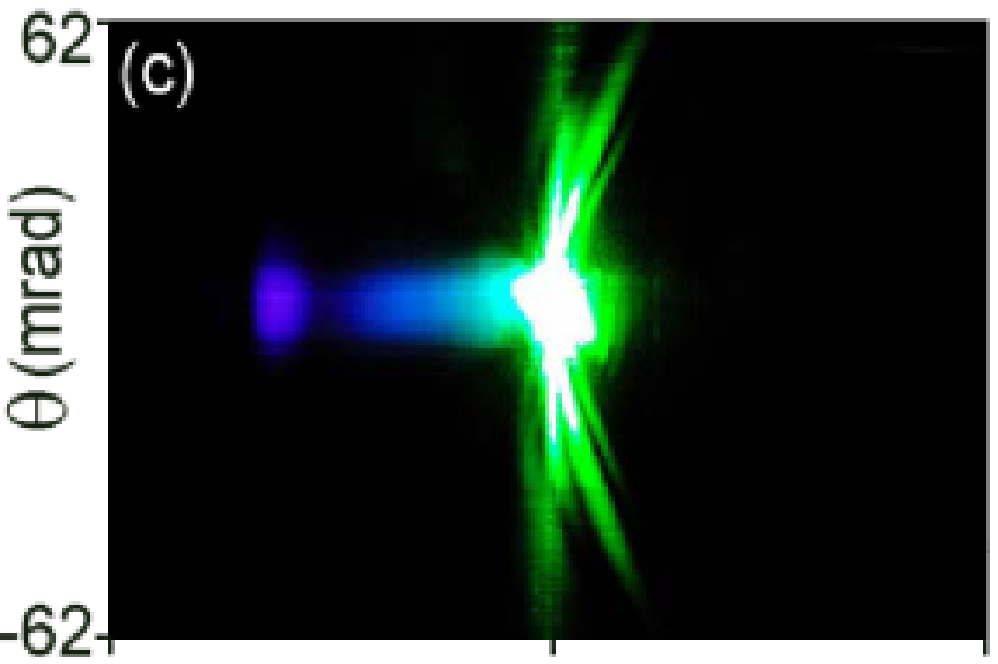}\\
\hspace*{0.1cm}\includegraphics[width=3.5cm]{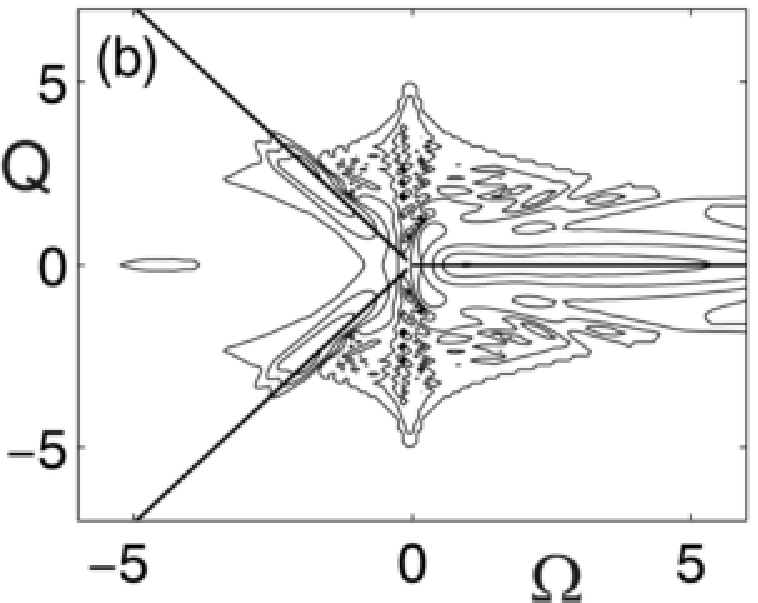}\hspace*{0.2cm}
\includegraphics[width=3.85cm]{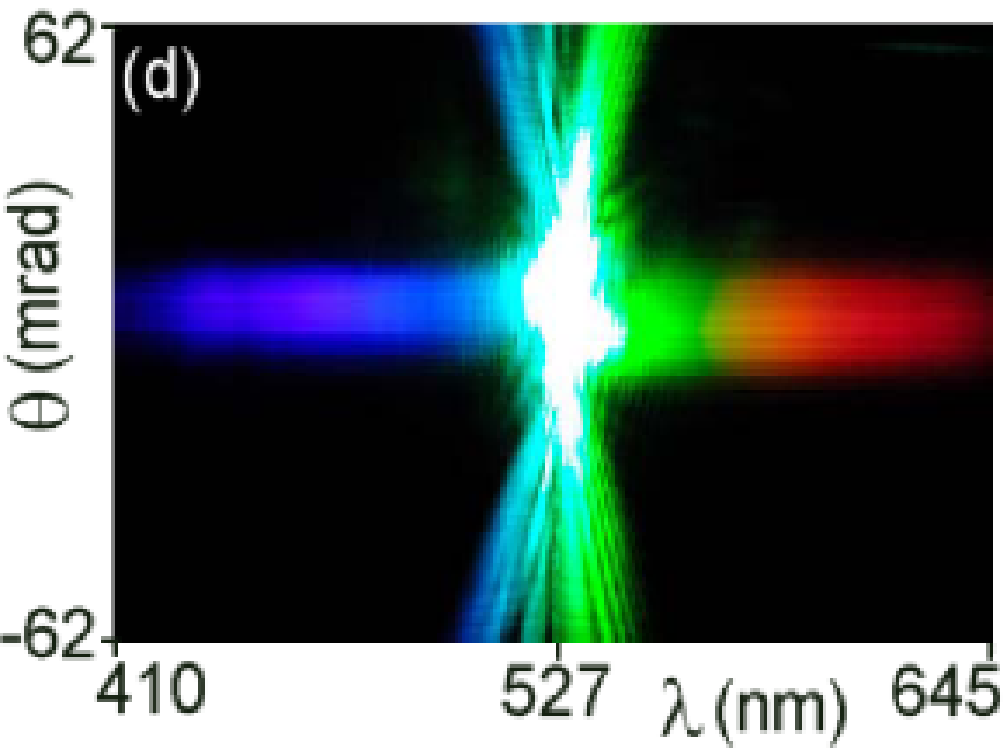}
\caption{\label{fig3} ST spectrum (a) of input asymmetric pulsed TP (see text),
(b) after propagation a short distance (subtracted of the input spectrum). Angularly resolved spectra of filaments in fused
silica at (c) $E=2\, \mu$J and (d) $E=3\, \mu$J.}
\end{figure}

It follows from our analysis that the onset of downshifted axial emission must be accompanied by upshifted CE, and vice versa,
and that these two events may occur independently. Our simulations and experiments support this interpretation. Figure
\ref{fig3}(a) shows the ST spectrum of a TP that is not a CW but a pulse. This ``perturbed" TP is moreover asymmetric in time by
choosing different durations for the leading and trailing parts ($\Delta\tau=0.87$ and $29$, respectively). In Fig.
\ref{fig3}(b) the difference between the propagated spectrum under the NSE (1) (distance $\zeta=7.6$) and the input spectrum is
shown in order to visualize the newly generated frequencies. A single Y-shaped structure is formed, with a slope of the conical
part fitting to $\sqrt{2}$ (thick line).

In the experiments, we used a 15 cm-long fused silica sample as Kerr media. 200-fs-long pulses at 527 nm delivered from a 10 Hz
Nd:glass mode-locked and regeneratively amplified system (Twinkle, Light Conversion), were spatially filtered and focused with a
50 cm focal length lens. The pulses then entered into the sample, whose input facet was placed at 52 cm from the lens, and
formed a single filament for input energies $E\gtrsim 2\,\mu$J. Angularly resolved spectra of the filament at the output facet
and from single laser shots were measured with an imaging spectrometer and a CCD camera, as described in \cite{FACCIO}. At 2
$\mu$J in fused silica the filament is formed just before the output facet of the sample, the Y-shaped (blue axial, red conical)
spectrum of Fig. \ref{fig3}(c) being then observed. At 3 $\mu$J, the filament is formed closer to the input facet. The double Y
spectrum of Fig. \ref{fig3}(d) then corresponds to a longer filament path within the sample. The faster growth of one of the two
unstable perturbations at the initial stage of filamentation supposes some unbalancing in their seeds. In fact, an expected
instability seeding mechanism, as self-phase modulation, generates preferentially axial, blue-shifted frequencies in the
presence of plasma and third-order dispersion.

ST instability of the TP also incorporates a mechanism of temporal splitting. If the perturbation $p_\kappa$ (e.g.) is seeded
coherently at different frequencies, its growth leads to the formation of a pulse, whose spectrum is increasingly peaked at the
maximum gain frequencies $\Omega_M\simeq 0.365$ for $u$ and $-\Omega_M\simeq -0.365$ for $v$. For $u$ and $v$ with respective
axial wave number shifts $\kappa_R+\alpha$ and $-\kappa_R+\alpha$, the inverse group velocity $1/v^{(g)}$ (in the frame moving
with the group velocity of a plane pulse at the TP frequency) will be $[d\kappa_R/d\Omega|_{\Omega_M}] \simeq 0.5$ for $u$, and
$[d(-\kappa_R)/d\Omega|_{-\Omega_M}]\simeq 0.5$ for $v$. Being equal, the Y-wave propagates as a whole with a well-defined group
velocity. For the $p_{-\kappa^\star}$ perturbation, or reflected Y-wave, the group velocity is the opposite, and the group
mismatch between the two Y-waves is $1/v^{g}_{p_\kappa} - 1/v^{(g)}_{p_{-\kappa^{\star}}}\simeq 1$, which in physical variables
yields, $\sqrt{k_0k_0^{\prime\prime}n_2 I/n_0}$, in agreement with the observed dependence in filamentation on pump intensity
and material properties \cite{FACCIO2}.

If the pump is not strictly monochromatic but a long pulse, the Y-waves will leave it at a certain stage, ceasing then to grow.
The $p_\kappa$ Y-wave behaves then as a linear wave of zero mean frequency and zero axial wave number shifts, since $u$ and $v$,
with opposite frequency shifts $\pm\Omega_M$, have also opposite axial wave number shifts
$\Omega_M^2/2-Q_u^2/2\simeq\Omega^2_M/2$ and $\Omega^2_M/2-Q_v^2/2\simeq -\Omega^2_M/2$. The peculiarity of the linear Y-wave is
that the $u$ part experiences the normal GVD $+\Omega^2/2$ of the medium, while the $v$ part experiences a net anomalous GVD
$-\Omega^2/2$ as a result of material and angular dispersion. This makes the $u$ and $v$ parts of $p_\kappa$ to continue to
depart from the pump at identical group velocity $[d(\Omega^2/2)/d\Omega|_{\Omega_M}] = [d(-\Omega^2/2)/d\Omega|_{-\Omega_M}]=
\Omega_M\simeq 0.365$ (again, the group velocity of the $p_{-\kappa^\star}$ Y-wave is the opposite).

Extending our analysis, we may venture an explanation to the fact that two X-waves are commonly observed {\em at later stages}
of propagation \cite{KOLESIK,FACCIO2}. Consider, within the FWM approach, the possible effects on the instability spectrum of
the strong {\em temporal localization} of the pump, as may take place upon (possibly multiple) splitting. For a spatially and
temporally localized pump of central frequency $\Omega=0$, axial wave vector shift $k_{\rm NL}$ due to SPM, and propagating at a
group velocity $v^{(g)}$ different from that of a plane pulse at $\Omega=0$ (as for an Y-wave), new plane waves $u$ and $v$ at
opposite frequencies $\pm\Omega$ are expected to be preferentially amplified if in addition to axial phase matching [Eq.
(\ref{MATCHING})], the velocity of the group formed by $u$ and $v$ matches the velocity of the pump. Equating the inverse beat
group velocity $[(\Omega^2/2 -Q_u^2/2) - (\Omega^2/2-Q_v^2/2)]/2\Omega$ between $u$ and $v$ to $1/v^{(g)}$, we obtain
$Q^2_v-Q_u^2 = 4\Omega/v^{(g)}$. Phase and group matching then yield $Q_{u,v}= \sqrt{\Omega^2 -(\pm\Omega)/v^{(g)} - 2k_{\rm
NL}}$ ($+\Omega$ for $u$, $-\Omega$ for $v$), which is the so-called dispersion curve of a frequency-gap X-wave mode
\cite{PORRAS,FACCIO2}. If this process takes place for the two Y-waves with group velocities $1/v^{(g)}=\pm\Omega_M$, the two
X-waves of Fig. \ref{fig2}(a, red and blue curves) with opposite frequency gaps are generated from the two Y-waves. On
nonlinearity relaxation ($k_{\rm NL}\rightarrow 0$) at increasing distances, one branch of each X-wave is seen to pass through
the point $Q=\Omega=0$ of the spectrum, as frequently observed \cite{KOLESIK,FACCIO2}.

In conclusion, Y-waves feature the MI of the TP and constitute the missing links between self-similar spatial self-focusing and
ST collapse driven dynamics. Y-waves model the coupling between axial and CE, and allow us to interpret temporal splitting,
X-wave formation and final relaxation into linear X-waves. These results provide a unified view of ultrashort pulse
filamentation and are relevant to all systems involving ST coupling and nonlinearity such as $\chi^{(2)}$ solitons, BEC, etc.

\end{document}